\documentclass[%
prl,
 amsmath,amssymb,
 aps,
floatfix,
]{revtex4-2}

\usepackage{graphicx}
\usepackage{dcolumn}
\usepackage{bm}
\usepackage{physics}

\newcommand{\Gr}{\mathsf{G}}

\usepackage{nicefrac}   
\newcommand{\half}[1]{\nicefrac{#1}{2}}

\newcommand{\nfrac}[2]{\nicefrac{#1}{#2}}
\newcommand{\Li}[1]{\text{Li}_{#1}}

\usepackage{color}

\begin{document}

\preprint{APS/123-QED}

\title{Information geometry and Bose-Einstein condensation}

\author{Pedro Pessoa}
 \affiliation{Physics Department, University at Albany (SUNY), Albany, NY 12222, USA.}
 \email{ppessoa@albany.edu}


\begin{abstract}
{
It is a long held conjecture in the connection between information geometry (IG) and thermodynamics that the curvature endowed by IG diverges at phase transitions. 
Recent work on the IG of Bose-Einstein (BE) gases challenged this conjecture by saying that in the limit of fugacity approaching unit --- where BE condensation is expected --- curvature does not diverge, rather it converges to zero. 
However, as the discontinuous behavior that identify condensation is only observed at the thermodynamic limit, a study of IG curvature at finite number of particles, $N$, is in order from which the thermodynamic behaviour can be observed by taking the thermodynamic limit ($N\to \infty$) posteriorly. This article presents such study, which was made possible by the recent advances presented in [Phys. Rev. A \textbf{104}, 043318 (2021)].
We find that for a trapped gas, as $N$ increases, the values of curvature decrease proportionally to a power of $N$ while the temperature at which the maximum value of curvature occurs approaches the usually defined critical temperature. 
This means that, in the thermodynamic limit, curvature has a limited value where a phase transition is observed, contradicting the forementioned conjecture.
}
\end{abstract}

\maketitle


\section{Introduction}

Information geometry (IG) \cite{Amari16,Ay17,Caticha15,Nielsen20}
is the study of the differential geometry structure of probability distributions. IG has found relevant results in several areas of science --- e.g. machine learning and data analysis \cite{Boso20,Dixit20}, stochastic thermodynamics \cite{Ito18}, quantum information processing \cite{Zanardi14,Banchi14}, economics and finance \cite{Brody01},  
and network sciences \cite{Costa21,Felice14}.
Meanwhile, Ruppeiner geometry \cite{Ruppeiner95} is an application of differential geometry to thermodynamics, developed as a geometric extension of Einstein's theory of fluctuations \cite{Ruppeiner79}, which was later found to be the IG of Gibbs distributions \cite{Brody95} --- {since Ruppeiner geometry does not connect directly to probabilities, it has attracted recent interest in the study of black hole thermodynamics \cite{Ruppeiner18,Wei19,Xu20,DiGennaro22}}.
One of the main claims of Ruppeiner geometry consists of a conjecture that the curvature induced from IG diverges at phase transitions \cite{Ruppeiner95,Ruppeiner10}, which are characterized as the non-analytical behaviour of some thermodynamic quantity in the thermodynamic limit -- the limit at which the number of particles $N$ goes to infinity, meant to represent that the number of particles in a gas is of the order of the Avogadro number $N_A \approx 6.02 \times 10^{23}$.

Bose-Einstein (BE) condensation is, arguably, the better studied example of phase transition --- and often the first to be taught in statistical physics textbooks (e.g. Landau\cite{Landau}).
In many thermodynamically relevant models, the gas consists of particles trapped in a potential that admits a density of states $\Gr$ as function of the state's energy $\epsilon$ of the form $\Gr(\epsilon) = \kappa \epsilon^\eta$. The exponent $\eta$ is given by the trapping potential. {For example, It can be found \cite{Aguilera-Navarro99,Pessoa21c}} that a harmonically trapped gas in $D$ dimensions yields $\eta = D-1$, and a gas in a regular box of $D$ dimensions implies $\eta = \half{D}-1$ --- and $\kappa$ has units of $[\text{energy}]^{-(\eta+1)}$. 
The grand canonical ensemble of a non-interacting gas of bosons yields the following relationship \footnote{The calculation leading to \eqref{NBE} can be seen in most, if not all, statistical physics textbooks e.g. \cite{Landau}, but is usually restricted to the gas trapped in a regular 3-dimensional box --- $\eta=\half{1}$. For a calculation for general density of states exponent, $\eta$, I refer to \cite{Aguilera-Navarro99} and \cite{Pessoa21c}} between the number of particles $N$, the temperature $T$, and the fugacity $\xi \in [0,1)$ :
\begin{equation}\label{NBE}
    N =  \kappa \frac{ \Gamma(\eta+1)}{\beta^{\eta+1}} \Li{\eta+1}( \xi)   + \frac{\xi}{1-\xi} \ ,
\end{equation}
where $\beta$ is the inverse temperature, $\beta = \frac{1}{k_{B}T}$ with $k_{B}$ being the  Boltzmann constant, and $\Li{}$ is the polylogarithm  \cite{wolframpages}
\begin{equation} \label{Polylogdef}
\Li{\varphi}(y) = \frac{1}{\Gamma(\varphi)} \int_0^\infty du \ \frac{u^{\varphi-1}}{y^{-1} e^u-1} = \sum_{k=1}^\infty \frac{y^k}{k^\varphi}\ ,
\end{equation}
with $\Gamma(\varphi)$ being the Euler's gamma function. The last term in \eqref{NBE} refers to the particles in the ground state. 
The critical inverse temperature $\beta_c$  for a BE gas is defined as
    \begin{equation}\label{critical}
        \beta_c \doteq \left[ \kappa \frac{\Gamma(\eta+1)}{N} \zeta(\eta+1) \right]^{\frac{1}{\eta+1}} \ , 
    \end{equation}
for $\eta>0$ \footnote{For $\eta\leq0$ $\beta_c$ diverges, so the critical temperature goes to the absolute zero.}, where $\zeta$ refers to the Riemann zeta function. In the thermodynamic limit the number of particles in the ground state goes to zero when $\beta<\beta_c$ and the fugacity $\xi$ approaches unit when $\beta \geq \beta_c$. 

Janyszek and Mrugala (JM) were the first, to the best of my knowledge, to study the IG of the a gas of bosons \cite{Janyszek90}. JM reported that the IG curvature diverges as $\xi \to 1^-$. This was received in the literature as a successful example of Ruppeiner's curvature conjecture \cite{Ruppeiner95,Oshima99,Mirza10}. However such study completely ignored the ground state term in \eqref{NBE}.  It was only recently \cite{Pessoa21,LopezPicon21} that JM results were challenged by observing that, if one keeps the ground state term in \eqref{NBE}, it follows that curvature does not diverge as $\xi \to 1^-$,  rather it converges to zero, in apparent contradiction to Ruppeiner's curvature conjecture. 
This was not the first criticism to Ruppeiner's curvature conjecture, Brody  and Rivier \cite{Brody95} reported that for the van der Waals model curvature diverges along the entire spinodal, not only at critical point, while Dey \emph{et al.} \cite{Dey12} report that curvature is smooth at the  Dicke model phase transition.

It must be noticed, however, that the BE phase transition is not identified directly by fugacity approaching unit, but rather by the fact that, in the thermodynamic limit, some thermodynamic relevant quantities --- namely the fraction of particles in the ground state for $\eta > 0$ and specific heat for $\eta \geq \half{1}$  are not analytical at the critical temperature \cite{Landau,Aguilera-Navarro99,Pessoa21c}. Hence, one might say that although \cite{Pessoa21,LopezPicon21} does correct the description given by JM \cite{Janyszek90}, it still does not provide a final test to Ruppeiner's conjecture. One ought to create a description in terms of $\beta$ and $N$ --- rather than $\beta$ and $\xi$ --- and the divergence  would appear, or not, at the critical temperature when one takes the thermodynamic limit.

The lack of a closed form expression for $\xi(\beta,N)$ --- defined as the inverse of \eqref{NBE} --- was the major difficulty preventing the feasibility of such study. However, a method to obtain $\xi(\beta,N)$ numerically was presented recently \cite{Pessoa21c}. With this tool, the calculations made in \cite{Pessoa21} can be extended to numerical calculations in terms of $\beta$ and $N$ and posteriorly plotted as a function of the reduced inverse temperature, $\gamma \doteq \frac{\beta-\beta_c}{\beta_c}$.

This article presents the results of this study. The following section defines the relevant variables and the numerical results in the form of graphs. The final section will discuss what these results mean for Ruppeiner's curvature conjecture.

\vspace{-.5cm}
\section{Results}
The IG curvature of a grand canonical BE gas is given by\cite{Pessoa21}
 \begin{equation}\label{curvaturebose}
     R(\beta,\xi)     = - \frac{\beta^{\eta+1}}{2\kappa}  \frac{\mathbb B(\xi,\eta) + \frac{\beta^{\eta+1}}{\kappa} \mathbb B_c(\xi,\eta) }{\left(\mathbb A(\xi,\eta)+  \frac{\beta^{\eta+1}}{\kappa} \mathbb A_c(\xi,\eta)\right)^2}    \ , 
\end{equation}
where
\begin{subequations}
\begin{align}
    \mathbb A(x,\varphi) &=
\det \left[\begin{matrix}
{\Gamma(\varphi+3)} \Li{\varphi+2}(x)& {\Gamma(\varphi+2)} \Li{\varphi+1}(x) \\
{\Gamma(\varphi+2)} \Li{\varphi+1}(x)  &  {\Gamma(\varphi+1)} \Li{\varphi}(x) 
\end{matrix}\right] \ , \\
    \mathbb A_c(x,\varphi) &=
\det \left[\begin{matrix}
{\Gamma(\varphi+3)} \Li{\varphi+2}(x) &0 \\
{\Gamma(\varphi+2)} \Li{\varphi+1}(x)  &  \frac{x}{(1-x)^2}
\end{matrix}\right] \ , \\
    \mathbb B(x,\varphi) &=
\det \left[ \begin{matrix}
{\Gamma(\varphi+3)} \Li{\varphi+2}(x)  & {\Gamma(\varphi+2)} \Li{\varphi+1}(x)  & {\Gamma(\varphi+1)} \Li{\varphi}(x)  \\ 
{\Gamma(\varphi+4)} \Li{\varphi+2}(x)  & {\Gamma(\varphi+3)} \Li{\varphi+1}(x)  & {\Gamma(\varphi+2)} \Li{\varphi}(x)  \\ 
{\Gamma(\varphi+3)} \Li{\varphi+1}(x)  & {\Gamma(\varphi+2)} \Li{\varphi}(x) & {\Gamma(\varphi+1)} \Li{\varphi-1}(x) 
\end{matrix} \right] \ , \\
    \mathbb B_c(x,\varphi) &= 
\det \left[ \begin{matrix}
{\Gamma(\varphi+3)} \Li{\varphi+2}(x)  & {\Gamma(\varphi+2)} \Li{\varphi+1}(x)  &\frac{x}{(1-x)^2} \\ 
{\Gamma(\varphi+4)} \Li{\varphi+2}(x) & {\Gamma(\varphi+3)} \Li{\varphi+1}(x)  &0 \\ 
{\Gamma(\varphi+3)} \Li{\varphi+1}(x)  & {\Gamma(\varphi+2)} \Li{\varphi}(x)  &\frac{x(x+1)}{(1-x)^3}
\end{matrix} \right] \ ;
\end{align}
\end{subequations}
note that $R$ is unitless.
As discussed, in order to create a valid comparison to critical temperature \eqref{critical} one has to calculate the thermodynamic quantities in terms of $\beta$ and $N$ rather than $\beta$ and $\xi$ as above. 
Let 
\begin{equation}
    R(\beta,N) \doteq R(\beta,\xi(\beta,N)) \ ,
\end{equation} 
where $\xi(\beta,N)$ is obtained as the inverse of \eqref{NBE} --- as mentioned, to the best of my knowledge, there is no closed form analytical expression for $\xi(\beta,N)$, nevertheless the code for calculating  $\xi(\beta,N)$ numerically is available in my GitHub repository \cite{github} and further explanation is given in \cite{Pessoa21c}.

\begin{figure*}
\centering
    \includegraphics[width=.9\textwidth]{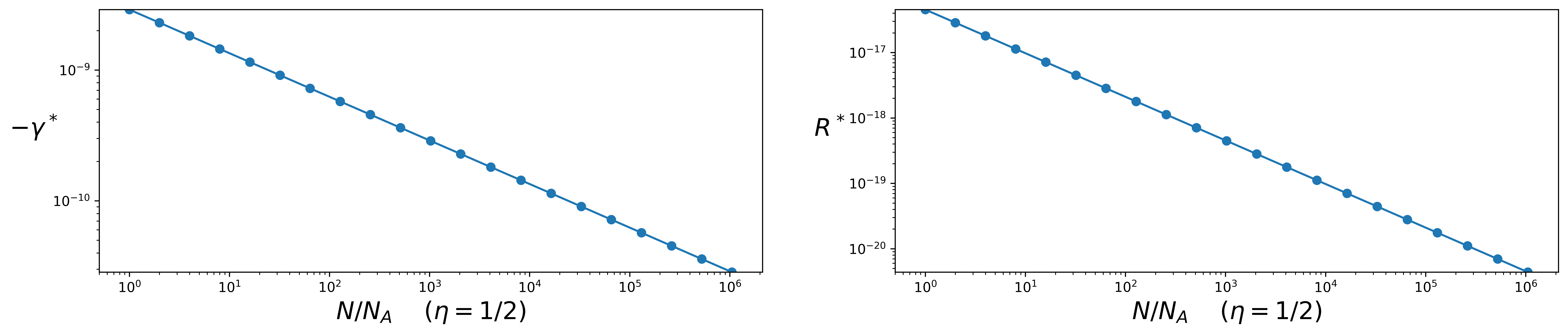}
    \includegraphics[width=.9\textwidth]{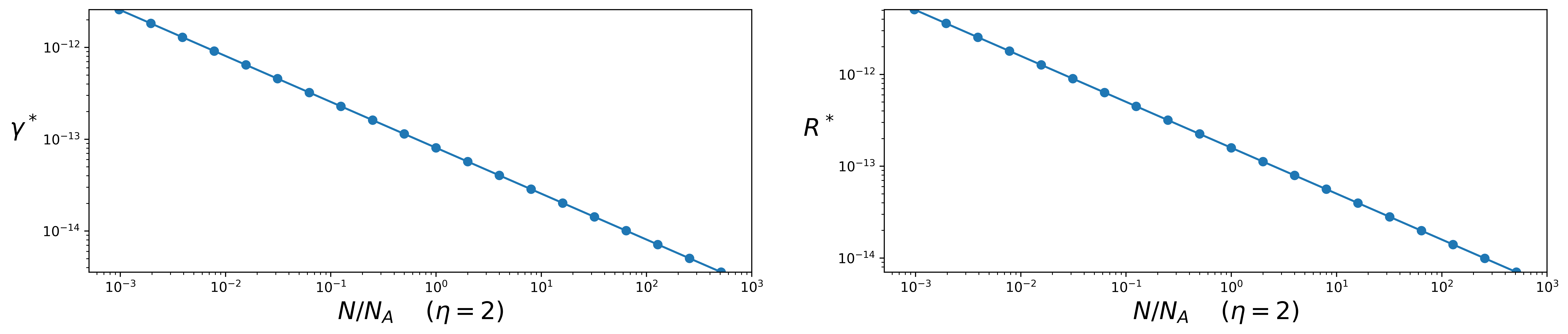}
    \vspace{-.4cm}
    \caption{Graphs for the numerically calculated maximum value of curvature, $R^*$, and the reduced inverse temperature at the maximum, $\gamma^*$, for different numbers of particles in order close to the one of the Avogadro number. The pair of graphs above corresponds to $\eta=\half{1}$ --- as in a gas trapped in a 3-dimensional box --- and the pair below corresponds to $\eta=2$ --- as in a 3-dimensional harmonically trapped gas. 
    Note that, as commented in the main text, for $\eta=\half{1}$ this yields the scaling relations $R^*(N) \approx 0.32339 N^{-\nfrac{2}{3}} $ and $\gamma^*(N) \approx -0.24473 N^{-\nfrac{1}{3}}$, while for $\eta=2$ it yields $R^*(N) \approx  0.12323N^{-\nfrac{1}{2}} $ and   $\gamma^*(N) \approx 0.062582 N^{-\nfrac{1}{2}}$. }
    \label{fig:Rstargamma}
\end{figure*}

In order to study how curvature behaves at phase transitions, one ought to define the quantities
\begin{subequations} \label{defs}
\begin{align}
    R^*(N) \doteq& \max_\beta R(\beta,N) \qq{and}\\
    \beta^*(N) \doteq&  \arg\max_\beta R(\beta,N)  \ . \label{defs_beta}
\end{align}
\end{subequations}
and observe how these behave for increasing $N$ \footnote{With the numerical calculation of $\xi(\beta,N)$ implemented in \cite{github}, it is possible to calculate $\beta^*$, and consequentially $R^*$, through the golden search method.}.
The results of numerical calculation for $R^*(N)$ and $\gamma^*(N) \doteq \frac{\beta^* - \beta_c}{\beta_c}$ for $\eta=\half{1}$ and $\eta=2$ are presented in Fig.\ref{fig:Rstargamma}. 
In it, one observes that while for increasing $N$ we have $\gamma^*(N)$ going to zero --- or $\beta^*$ approaching $\beta_c$ --- the value of $R^*$ decreases with increasing $N$.  
Roughly, for $\eta = \half{1}$, these numerical results suggest a large $N$ scaling of the form 
$R^*(N) \approx 0.32339 N^{-\nfrac{2}{3}} $ and 
$\gamma^*(N) \approx -0.24473 N^{-\nfrac{1}{3}}$; 
while for $\eta = 2$, these results suggest a large $N$ scaling of the form 
$R^*(N) \approx  0.12323N^{-\nfrac{1}{2}} $ and $\gamma^*(N) \approx 0.062582 N^{-\nfrac{1}{2}}$.

\begin{figure*}
\centering
    \vspace{-.25cm}
    \includegraphics[width=.45\textwidth]{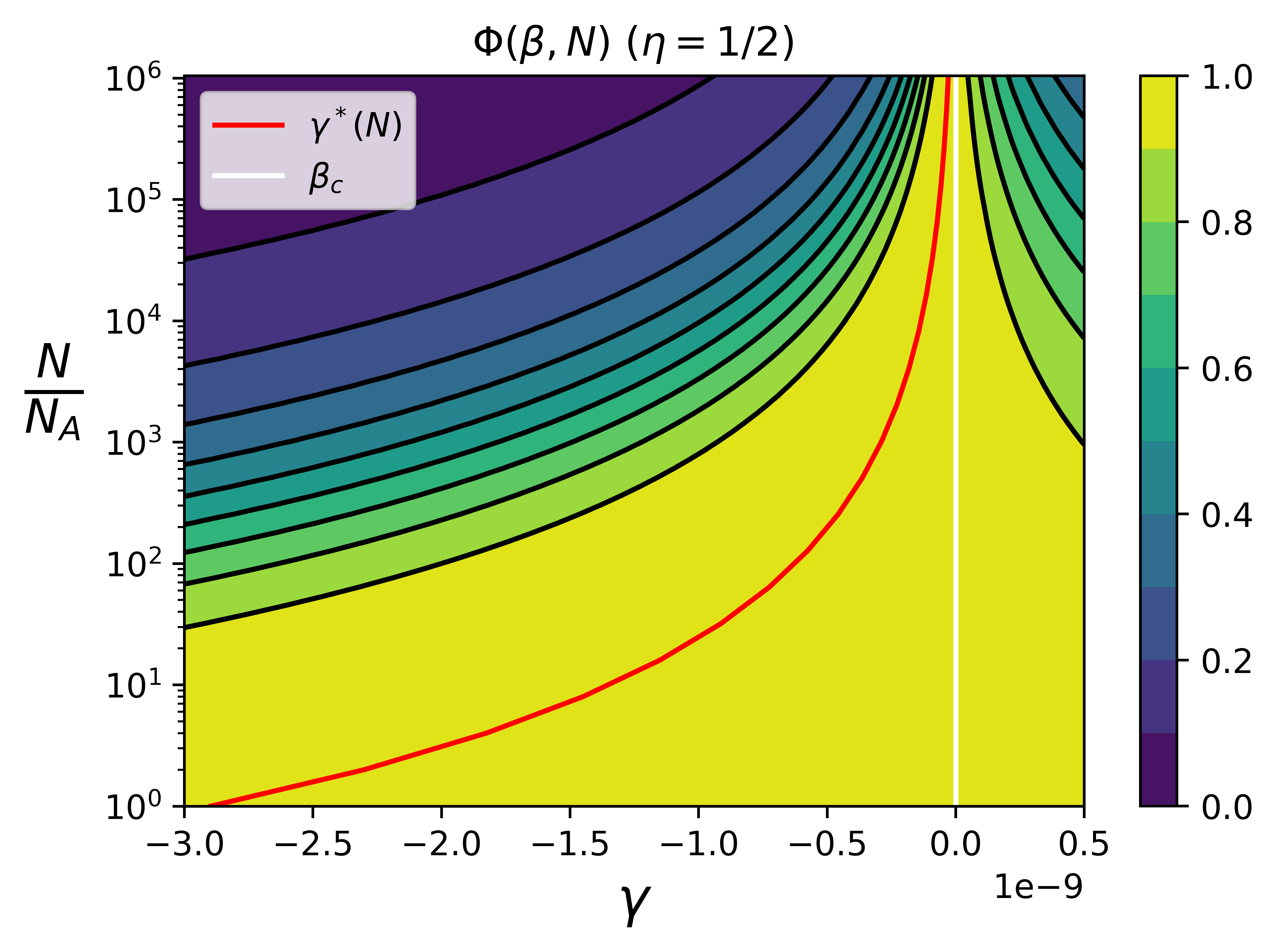}
    \hspace{.05\textwidth}
    \includegraphics[width=.45\textwidth]{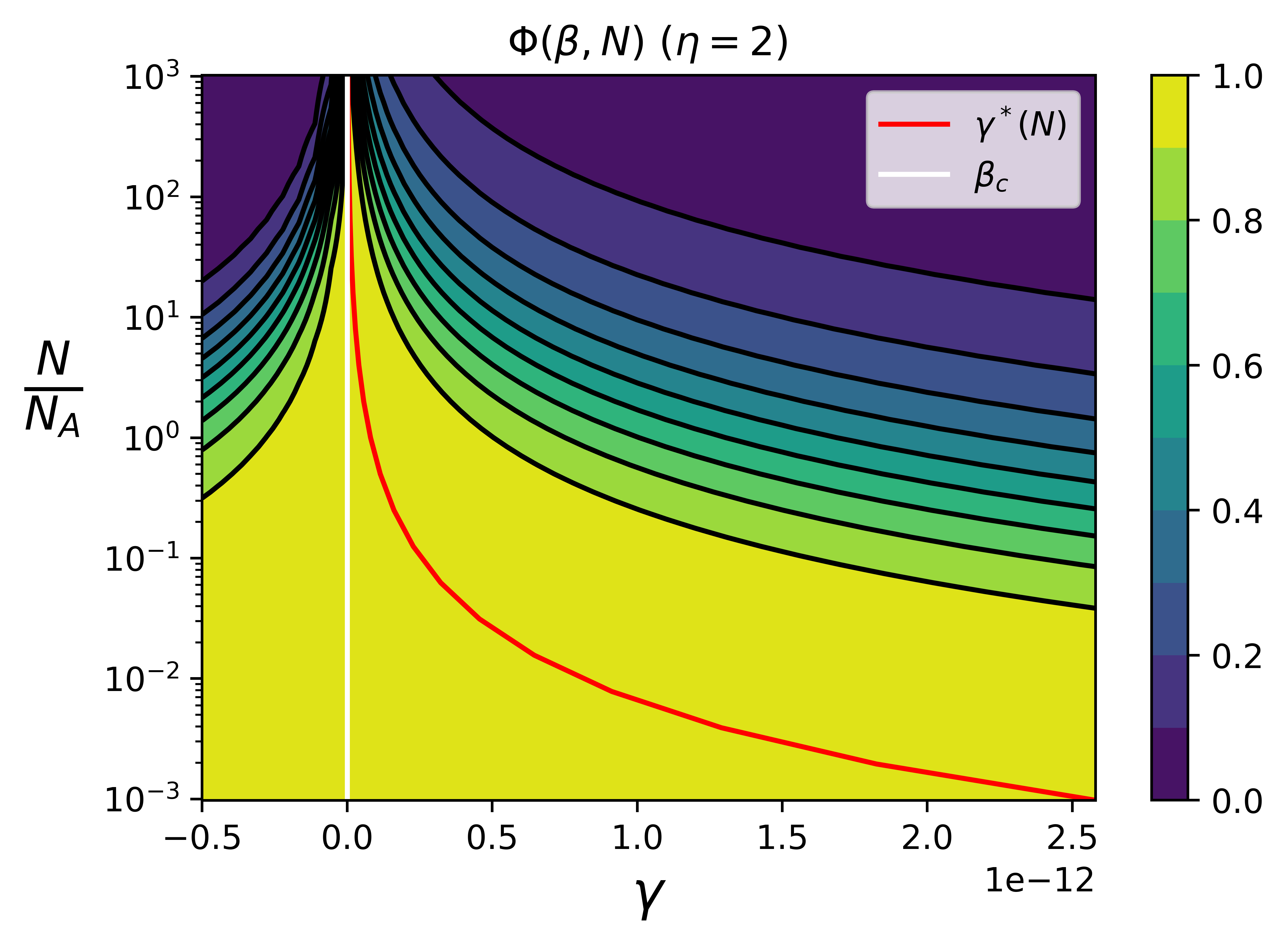}
    \vspace{-.4cm}
    \caption{Contour graphs for the curvature ratio, $\Phi(\beta,N) = \frac{R(\beta,N)}{ R^*(N) }$, in terms of $N$ and $\gamma = \frac{\beta - \beta_c}{\beta_c}$ for $\eta=\half{1}$ (on the left) and $\eta=2$ (on the right). 
    On both it is possible to see that curvature spikes in a small range of $\gamma$ values, which include the critical inverse temperature $\beta_c$ (equivalent to $\gamma=0$ represented above in white) and the reduced inverse temperature of the curvature maximum ($\gamma^* = \frac{\beta^* - \beta_c}{\beta_c}$ with $\beta^*$ defined in \eqref{defs_beta} in red).  The region where the curvature ratio is not negligible reduces as $N$ increases, indicating a large $N$ behaviour that would converge to $\Phi=0$ for $\beta \neq \beta_c$ and $\Phi=1$ at $\beta=\beta_c$. It can also be seen that, in accordance to Fig. \ref{fig:Rstargamma}, $\gamma^*$ converges to zero as $N$ increases.}
    \label{fig:Rrate}
    \vspace{-.25cm}
\end{figure*}

For an extra visualization of the behaviour of curvature at a finite number of particles, it is interesting to define the curvature ratio 
\begin{equation}
    \Phi(\beta,N) \doteq \frac{R(\beta,N)}{ R^*(N) }
\end{equation}
which measures how the curvature deviates from the maximum with inverse temperature. In Fig. \ref{fig:Rrate} we present the values of $\Phi$ for $\eta=\half{1}$ -- with $N$ ranging from $N_A$ to $2^{20} N_A$ --  and $\eta=2$ -- with $N$ ranging from $2^{-10} N_A$ to $2^{10} N_A$.  
One can note that, although the value of curvature decreases with increasing $N$  as seen in Fig. \ref{fig:Rstargamma}, the curvature ratio, $\Phi$, has a ``bandwidth'' around the maximum that reduces as $N$ increases. This suggests a sharp peak at $\beta_c$ in the thermodynamic limit. 

\vspace{-.5cm}
\section{Discussion}

The result presented in Fig. \ref{fig:Rstargamma} directly contradicts Ruppeiner's curvature conjecture for the BE condensation. If curvature diverged at the phase transition, one would see that $R(\beta_c,N)$ diverges at the thermodynamic limit. This is not possible as the maximum value of curvature reduces for increasing $N$. Writing curvature in terms of $\beta$ and $N$ is far more convincing what was previously reported \cite{Pessoa21,LopezPicon21} in studies in terms of $\beta$ and $\xi$ since it takes into account the fact that the non analytical behaviour identifying phase transitions only appear in the thermodynamic limit. 

On the other hand, one consequence of thermodynamic curvature diverging only at the critical temperature would be that $\Phi(\beta,N)$ converges, as $N\to\infty$, to zero anywhere but the critical temperature. This is consistent with the results presented in Fig. \ref{fig:Rrate}.
This means that, although BE condensation is a counter example of Ruppeiner's curvature conjecture, information geometry curvature explicitly distinguishes the BE critical point. 

\section*{Acknowledgments}
I would like to thank A. Caticha, D. Robbins and C. Cafaro for insightful discussions leading to this study.


\bibliography{ref}

\end{document}